\documentclass[hyper]{JHEP} % 10pt is ignored!

\usepackage{epsfig}
\newcommand\fverb{\setbox\pippobox=\hbox\bgroup\verb}

\newcommand\fverbdo{\egroup\medskip\noindent%

            \fbox{\unhbox\pippobox}\ }

\newcommand\fverbit{\egroup\item[\fbox{\unhbox\pippobox}]}

\newbox\pippobox

\title{Canonical Analysis of Unimodular Gravity}
\author{J. Kluso\v{n}\\
Department of
Theoretical Physics and Astrophysics\\
Faculty of Science, Masaryk University\\
Kotl\'{a}\v{r}sk\'{a} 2, 611 37, Brno\\
Czech Republic\\
E-mail: \email{klu@physics.muni.cz}}
\preprint{}

 \abstract{This short note is devoted to the Hamiltonian
 analysis of the Unimodular Gravity.We treat
 the unimodular gravity as General Relativity action with
the unimodular constraint imposed with the help of Lagrange
multiplier. We perform the canonical analysis of the resulting
theory and  determine its  constraint structure. }
 \keywords{Hamiltonian Formalism, Unimodular Gravity }

\def\be{\begin{equation}}

\def\ee{\end{equation}}

\def\bea{\begin{eqnarray}}

\def\eea{\end{eqnarray}}
\def\bmH{\bar{\mH}}

\def\tmH{\tilde{\mH}}

\def\mH{\mathcal{H}}

\def\bx{\mathbf{x}}
\def\by{\mathbf{y}}

\newcommand{\hg}{\hat{g}}

\newcommand{\mF}{\mathcal{F}}

\newcommand{\mG}{\mathcal{G}}

\newcommand{\bT}{\mathbf{T}}

\def\pb #1{\left\{#1\right\}}

\begin{document}
%%%%%%%%%%%%%%%%%%%%%
%%%%Introduction %%%%%%%%%
%%%%%%%%%%%%%%%%%%%%
\section{Introduction and Summary}\label{first}
Unimodular gravity is obtained from Einstein-Hilbert action in which
the unimodular condition
\begin{equation}\label{unicon}
\sqrt{-\det \hg_{\mu\nu}}=1
\end{equation}
is imposed from the beginning
\cite{Buchmuller:1988wx,Buchmuller:1988yn}. The resulting field
equations correspond to the traceless Einstein equations and can be
shown that they are equivalent to the full Einstein equations with
the cosmological constant term $\Lambda$, where $\Lambda$ enters as
an integration constant. In other words we see clear equivalence
between unimodular gravity and general relativity. On the other hand
the idea that the cosmological constant arises as an integration
constant is very attractive and it is one of the motivation for the
study of the unimodular gravity, for recent study, see
\cite{Ellis:2010uc,Ellis:2013uxa,Padilla:2014yea,Gao:2014nia,
Jain:2012gc,Smolin:2010iq,Smolin:2009ti,Shaposhnikov:2008xb,
Alvarez:2005iy,Finkelstein:2000pg,Barcelo:2014mua,Barcelo:2014qva}. 

The fact that the determinant of the metric is fixed has clearly
profound consequences on the structure of given theory. First of all
it reduces the full group of diffeomorphism to invariance under the
group of unimodular general coordinate transformations which are
transformations that leave the determinant of the metric unchanged.
Further, the fact that the metric is fixed could have important
consequences for the Hamiltonian formulation of given theory. Some
aspects of the Hamiltonian treatment of unimodular gravity were
analyzed in \cite{Unruh:1988in,Unruh:1989db}. Then very important
contribution to this analysis was presented in
\cite{Henneaux:1989zc}, where the condition (\ref{unicon}) was fixed
by hand from the beginning. On the other hand we mean that it would
be desirable to impose this condition using the Lagrange multiplier
term that is added to the gravity action. In fact, similar analysis
was performed in \cite{Kuchar:1991xd} using very elegant formalism
of geometrodynamics \cite{Isham:1984rz,Isham:1984sb} which is
manifestly diffeomorphism invariant. However this elegant
formulation can be achieved with the help of the introducing of the
collection of the scalar fields which on the other hand makes the
analysis more complicated.  Our goal is to perform the Hamiltonian
analysis in more straightforward manner when we consider general
relativity action where the constraint (\ref{unicon}) is imposed
with the help of the Lagrange multiplicator. Clearly this expression
 breaks the diffeomorphism invariance explicitly and we would like to see  the
consequence of the presence of this term on the Hamiltonian
structure of given theory. It turns out that given structure is
rather interesting. Explicitly, we consider Lagrange multiplicator
as the dynamical variable where its momentum is the primary
constraint of the theory. We also find that the momentum conjugate
to the lapse $N$ is not the first class constraint but together with
(\ref{unicon}) form the collection of the second class constraints.
Then we find another set of constraints that implies that the
Lagrange multiplier has to depend on time only. Finally we split the
Hamiltonian constraints into collection of $\infty^3-1$ constraints
(in terminology of \cite{Kuchar:1991xd}) and one constraint that
together with the momentum conjugate to the  zero mode part of the
Lagrange multiplier forms the second class constraints. This is
subtle difference with respect to the case of general relativity
that possesses $4\infty^3$ first class constraints.  On the other
hand the presence of the global constraint that relates the
dynamical gravity fields and embedding fields was mentioned in
\cite{Kuchar:1991xd} and we mean that our result has closed overlap
with the conclusion derived there.

As the next step we perform the Hamiltonian analysis of the
unimodular theory proposed in  \cite{Henneaux:1989zc}. Now due to
the fact that given theory is manifestly covariant the analysis is
more straightforward and we derive $4\infty^3$ first class
constraints. On the other hand the structure of the Hamiltonian
constraint is different from the Hamiltonian constraint of the
general relativity since now it contains the term corresponding the
momentum conjugate to time component of the vector field $\mF^\mu$.
Now due to the fact that the Hamiltonian does not depend on this
field explicitly we find that this momentum is constant on shell and
hence its constant value can be considered as an effective
cosmological constant.

%
%Let us outline our results. We perform the Hamiltonian analysis of
%the unimodular gravity and we find that it has almost the same
%structure as general relativity action with the subtle difference
%in the absence of one local Hamiltonian constraint. However we mean
%that this fact does not have an impact on the number of the local
%degrees of freedom.

This paper is organized as follows. In the next section
(\ref{second}) we perform the Hamiltonian analysis of unimodular
theory with constraint (\ref{unicon}) included into the action using
the Lagrange multiplicator. Then in section (\ref{third}) we perform
the Hamiltonian analysis of the formulation of unimodular gravity
proposed in \cite{Henneaux:1989zc}.

\section{ Hamiltonian Analysis of Unimodular Gravity }
\label{second} In this section we perform the Hamiltonian analysis
of the unimodular gravity  where the condition (\ref{unicon}) is
imposed using the Lagrange multiplier term included into the action.
Explicitly, we consider the action
\begin{eqnarray}\label{uniact}
S=\frac{1}{16\pi G}\int d^4x( \sqrt{-\hg}{}^{(4)}R[\hg]-\Lambda
(\sqrt{-\hg}-1)) \ ,
\nonumber \\
\end{eqnarray}
where ${}^{(4)}R$ is four dimensional curvature and where
$\Lambda(x) $ is Lagrange multiplicator.

To proceed to the canonical formulation we use the well know $3+1$
formalism that is the fundamental ingredient of the Hamiltonian
formalism of any theory of gravity \footnote{For recent review, see
\cite{Gourgoulhon:2007ue}.}. We consider $3+1$ dimensional manifold
$\mathcal{M}$ with the coordinates $x^\mu \ , \mu=0,\dots,3$ and
where $x^\mu=(t,\bx) \ , \bx=(x^1,x^2,x^3)$. We presume that this
space-time is endowed with the metric $\hat{g}_{\mu\nu}(x^\rho)$
with signature $(-,+,+,+)$. Suppose that $ \mathcal{M}$ can be
foliated by a family of space-like surfaces $\Sigma$ defined by
$t=x^0$. Let $g_{ij}, i,j=1,2,3$ denotes the metric on $\Sigma$ with
inverse $g^{ij}$ so that $g_{ij}g^{jk}= \delta_i^k$. We further
introduce the operator $\nabla_i$ that is covariant derivative
defined with the metric $g_{ij}$.
% The
% basic vector fields on $\Sigma_t$ are
% $\partial_i$.
% We  introduce  the
%future-pointing unit normal vector $n^\mu$ to the surface
%$\Sigma_t$. In ADM variables we have $n^0=\sqrt{-\hat{g}^{00}},
%n^i=-\hat{g}^{0i}/\sqrt{-\hat{g}^{ 00}}$.
 We also define  the lapse
function $N=1/\sqrt{-\hat{g}^{00}}$ and the shift function
$N^i=-\hat{g}^{0i}/\hat{g}^{00}$. In terms of these variables we
write the components of the metric $\hat{g}_{\mu\nu}$ as
\begin{eqnarray}
\hat{g}_{00}=-N^2+N_i g^{ij}N_j \ , \quad \hat{g}_{0i}=N_i \ , \quad
\hat{g}_{ij}=g_{ij} \ ,
\nonumber \\
\hat{g}^{00}=-\frac{1}{N^2} \ , \quad \hat{g}^{0i}=\frac{N^i}{N^2} \
, \quad \hat{g}^{ij}=g^{ij}-\frac{N^i N^j}{N^2} \ .
\nonumber \\
\end{eqnarray}
Then the standard canonical analysis leads to the bare Hamiltonian
in the form
\begin{equation}\label{Hbare}
H=\int d^3\bx (N\mH_T+N^i\mH_i+\Omega (\sqrt{g}N-1)+ v_N\pi_N+
v_i\pi_i+v_\Omega p_\Omega) \ ,
\end{equation}
where
\begin{eqnarray}
\mH_T= \frac{16\pi G}{\sqrt{g}} \pi^{ij}\mG_{ijkl}\pi^{kl}
-\frac{\sqrt{g}}{16\pi G} R \ , \quad \mH_i=-2g_{ik}\nabla_j
\pi^{jk} \  ,
\end{eqnarray}
where
\begin{equation}
\mG_{ijkl}=\frac{1}{2}(g_{ik}g_{jl}+g_{il}g_{jk})-
\frac{1}{2}g_{ij}g_{kl} \ ,
\end{equation}
and where $R$ is three dimensional curvature. Further, $\pi^{ij}$
are momenta conjugate to $g_{ij}$ with non-zero Poisson bracket
\begin{equation}
\pb{g_{ij}(\bx),\pi^{kl}(\by)}=\frac{1}{2}\left(\delta_i^k\delta_j^l+
\delta_i^l\delta_j^k\right)\delta(\bx-\by) \ .
\end{equation}
Finally, $\pi_N\approx 0 , \quad \pi_i\approx 0,\quad
p_\Omega\approx 0$ are primary constraints where
$\pi_N,\pi_i,p_\Omega$ are momenta conjugate to $N,N^i$ and $\Omega$
respectively with non-zero Poisson brackets
\begin{equation}
\pb{N(\bx),\pi_N(\by)}=\delta(\bx-\by) \ , \quad \pb{N^i(\bx),
\pi_j(\by)}=\delta^i_j\delta(\bx-\by) \ , \quad
\pb{\Omega(\bx),p_\Omega(\by)}=\delta(\bx-\by) \ .
\end{equation}
It is also useful to introduce the smeared form of the constraints
$\mH_T,\mH_i$
\begin{equation}
\bT_T(X)=\int d^3\bx X\mH_T \ , \quad \bT_S(X^i)=\int d^3\bx
X^i\mH_i \ ,
\end{equation}
where $X,X^i$ are functions on $\Sigma$. For further purposes we
also introduce the well known Poisson brackets
\begin{eqnarray}\label{pbbTS}
\pb{\bT_T(X),\bT_T(Y)}&=&\bT_S((X\partial_i Y-Y\partial_i X)g^{ij})
\
, \nonumber \\
\pb{\bT_S(X),\bT_T(Y)}&=&\bT_T(X^i\partial_iY) \ , \nonumber \\
\pb{\bT_S(X^i),\bT_S(Y^j)}&=&\bT_S(X^j\partial_j Y^i- Y^j\partial_j
X^i) \ . \nonumber \\
\end{eqnarray}
Now we proceed to the analysis of the preservation of the primary
constraints. Explicitly from (\ref{Hbare})  we find
\begin{eqnarray}
\partial_t\pi_N&=&\pb{\pi_N,H}=-\mH_T-\sqrt{g} \Omega \equiv
-\mH'_T\approx 0 \ , \nonumber \\
\partial_t p_\Omega&=&\pb{p_\Omega,H}=-(N
\sqrt{g}-1)\equiv -\Gamma \approx 0 \ , \nonumber \\
\partial_t\pi_i&=&\pb{\pi_i,H}=-\mH_i \approx 0 \ . \nonumber \\
\end{eqnarray}
 Then the total Hamiltonian with all constraints included has the
form
\begin{eqnarray}\label{HT}
H_T=\int d^3\bx (N\mH_T+v_T\mH_T'+(v_\Gamma+\Omega) \Gamma+
N^i\mH_i+v_\Omega p_\Omega+v_N\pi_N) \ , \nonumber \\
\end{eqnarray}
where $ N^i$ can now be considered as Lagrange multipliers
corresponding to the constraints $ \mH_i$ while we still keep $N$ as
dynamical variable while $v_T$ and $v_\Omega$ are  the Lagrange
multipliers corresponding to the constraints $\mH'_T$ and $\Gamma$
respectively.

Now we proceed to the analysis of the stability of all constraints.
Using (\ref{HT}) we find
\begin{eqnarray}
\partial_t p_\Omega=\pb{p_\Omega,H_T}=-\Gamma-v_T \sqrt{g}
\approx -v_T\sqrt{g} \nonumber \\
\end{eqnarray}
that implies $v_T=0$. In case of the constraint $\pi_N\approx 0$ we
have
\begin{equation}
\partial_t\pi_N=\pb{\pi_N,H_T}=-\mH_T -(v_\Gamma+\Omega)\sqrt{g}
=-\mH_T'-v_\Gamma\sqrt{g}=0
\end{equation}
that again implies that $v_\Gamma=0$. Let us now consider the time
evolution of the constraint $\Gamma$
\begin{eqnarray}
\partial_t \Gamma&=&\pb{\Gamma,H_T}=\nonumber \\
&\equiv & -8\pi G N^2 \pi^{ij}g_{ij}+\partial_i N^i N\sqrt{g}+
v_N\sqrt{g}
=0\nonumber \\
\end{eqnarray}
that can be considered as the equation for the Lagrange multiplier
$v_N$. In case of  the constraint $\mH'_T$ we find
\begin{eqnarray}\label{partmH}
\partial_t \mH'_T=\pb{\mH'_T,H_T}=
\int d^3\bx (\pb{\mH'_T,N\mH'_T}+
\sqrt{g}v_\Omega\approx \sqrt{g}v_\Omega=0 \ , \nonumber \\
\end{eqnarray}
where in the first step we used (\ref{pbbTS}). Then (\ref{partmH})
implies  $v_\Omega=0$.  Finally we consider the time evolution of
the constraint $\mH_i$. Due to the fact that $\Omega$ and $N$ are
dynamical variables  it is natural to extend the constraint $\mH_i$
with the appropriate combination of the primary constraints
$p_\Omega$ and $\pi_N$ so that
\begin{equation}
\tmH_i=\mH_i+p_\Omega\partial_i\Omega+\pi_N\partial_i N \ ,
\bT_S(N^i)= \int d^3\bx N^i\tmH_i \ .
\end{equation}
Now the time evolution of the smeared form of the constraint
$\bT_S(M^i)$ is equal to
\begin{eqnarray}\label{partbTS}
\partial_t\bT_S(M^i)&=&\pb{\bT_S(M^i),H_T}\approx
\pb{\bT_S(M^i),\int d^3\bx N\mH'_T}-\pb{\bT_S(M^i),\Omega}\approx \nonumber \\
&\approx & M^i\partial_i \Omega =0 \ ,   \nonumber \\
\end{eqnarray}
where we again used (\ref{pbbTS}). Since the equation above has to
be valid for all $M^i$ we see that it corresponds to some form of
the constraint on $\Omega$. In order to explicitly identify the
nature of given constraint we split $\Omega$ into the zero mode part
and the remaining part as follows
\begin{equation}
\Omega(\bx,t)=\Omega_0(t)+\bar{\Omega}(\bx,t) \ , \quad
\Omega(t)=\frac{1}{\int d^3\bx\sqrt{g}}\int d^3\bx \sqrt{g}
\Omega(\bx,t) \ ,
\end{equation}
where by definition $\int d^3\bx \sqrt{g}\bar{\Omega}(\bx,t)=0$.
Then the equation (\ref{partbTS}) implies
\begin{equation}
\bar{\Omega}(\bx,t)=K(t) \ ,
\end{equation}
where from definition of $\bar{\Omega}$ we obtain
\begin{equation}
\int d^3\bx \sqrt{g}\bar{\Omega}(\bx,t)= K(t)\int d^3\bx \sqrt{g}=0
\end{equation}
and hence we find $K(t)=0$. In other words we have following
 constraint
\begin{equation}
\bar{\Omega}(\bx,t)=0 \
\end{equation}
while the zero mode $\Omega_0(t)$ is still non-specified. It is
useful to perform the similar separation of the zero mode part of
$p_\Omega$ as well
\begin{eqnarray}
p_\Omega(\bx,t)&=&\frac{\sqrt{g}}{\int d^3\bx \sqrt{g} }
P_\Omega(t)+\bar{p}_\Omega(\bx,t) \ , \nonumber \\
p_\Omega(t)&=&\int d^3\bx p_\Omega(\bx,t) \ , \quad  \int d^3\bx
\bar{p}_\Omega(\bx,t)=0 \ . \nonumber \\
\end{eqnarray}
Note that we included the factor $\frac{\sqrt{g}}{\int d^3\bx
\sqrt{g}}$ in front of $p_\Omega$ in order to have canonical Poisson
bracket
\begin{equation}
\pb{\Omega_0,P_\Omega}=1
%\frac{1}{\int d^3\bx \sqrt{g}} \int d^3\bx
%d^3\by \sqrt{g}(\bx)\pb{\Omega(\bx),p_\Omega(\by)}= 1 \
\end{equation}
and also in order to ensure that $p_\Omega$ transforms as density
since $P_\Omega$ is scalar.
 Then by definition we also  find
%\begin{eqnarray}
%\pb{\Omega(\bx),P_\Omega}= \int d^3\by
%\pb{\Omega(\bx),p_\Omega(\by)}=\int d^3\by \delta(\bx-\by)=1 \ ,
%\nonumber \\
%\pb{p_\Omega(\bx),\Omega_0}= \frac{1}{\int d^3\bx\sqrt{g}} \int
%d^3\by \sqrt{g}(\by)\pb{p_\Omega(\bx),\Lambda(\by)}=-\frac{\sqrt{g}}{\int d^3\bx
%\sqrt{g}}\nonumber \\
%\end{eqnarray}
%and hence
%\begin{eqnarray}
%\pb{\bar{\Omega}(\bx),\bar{p}_\Omega(\by)}=
%\pb{\Omega(\bx)-\Omega_0,p_\Omega(\by)-\frac{\sqrt{g}}{\int
%d^3\bx\sqrt{g}}P_\Omega}=\delta(\bx-\by)-\frac{\sqrt{g}(\by)}
%{\int d^3\bx \sqrt{g}} \ . \nonumber \\
%\end{eqnarray}
%and also
\begin{eqnarray}
\pb{\bar{\Omega}(\bx),P_\Omega}=0 \ , \pb{\bar{p}_\Omega,\Omega_0}=0
\ . \nonumber \\
\end{eqnarray}
%Let us now outline number of the constraints
%\begin{eqnarray}
%\pi_N\approx 0 \ , P_\Omega \approx 0  \ , \bar{p}_\Omega\approx 0 \
%, \bar{\Omega}\approx 0 \ , \Gamma:\sqrt{g}N-1\approx 0 \ ,
% \mH'_T=\mH_T+\sqrt{g}\Lambda_0 \ . \nonumber \\
% \end{eqnarray}
%To proceed further it is useful to
 It turns out that it is useful
to perform similar separation in case of the constraint $\mH_T$
\begin{equation}\label{sepmHT}
\mH_T=\frac{\sqrt{g}}{\int d^3\bx\sqrt{g}} \mH_0+\bar{\mH}_T \ ,
\quad \mH_0=\int d^3\bx \mH_T \ , \quad \int d^3\bx\bar{\mH}_T=0 \
\end{equation}
and also in case of  the Lagrange multiplicator $v_N$
\begin{equation}\label{sepvN}
v_N=v^N_0+\bar{v}_N \ , \quad  v_0^N=\frac{1}{\int
d^3\bx\sqrt{g}}\int d^3\bx \sqrt{g}v_N \ , \quad  \int d^3\bx
\sqrt{g}\bar{v}_N=0 \ .
\end{equation}
%and hence
%\begin{eqnarray}
%\int d^3\bx v_N\mH_T=v_0^N\mH_0+\int d^3\bx \bar{v}_N
%\bar{\mH}_T \ , \nonumber \\
%\int d^3\bx v_N \Omega_0\sqrt{g}=\Omega_0 \int d^3\bx \sqrt{g} \ .
%\nonumber \\
%\end{eqnarray}
Note that the Poisson brackets between $\bar{\mH}_T$ still have the
form as (\ref{pbbTS}). Explicitly, let us  define smeared form of
this constraint
\begin{equation}
\bar{\bT}_T(N)=\int d^3\bx N(\bx)\bmH_T(\bx)= \int d^3\bx
\bar{N}(\bx)\bmH_T(\bx) \ ,
\end{equation}
where we performed the separation $N=N_0+\bar{N} \ , \int d^3\bx
N\sqrt{g}=0$. Then we have
\begin{eqnarray}\label{pbbTSbar}
\pb{\bar{\bT}_T(N),\bar{\bT}(M)}=
\pb{\bT_T(\bar{N}),\bT_T(\bar{M})}=\int d^3\bx ((\bar{N}\partial_i
\bar{M}-\partial_i\bar{N} \bar{M})g^{ij}\mH_j)\  \nonumber \\
\end{eqnarray}
and hence the right side vanishes on the constraint surface
$\mH_i\approx 0$.  With the help of the separation (\ref{sepmHT})
and (\ref{sepvN}) we find the total Hamiltonian in the form
\begin{equation}
H_T=\int d^3\bx (N\mH_T+\bar{v}_N\bar{\mH}_T+v^N_0 \int d^3\bx
\sqrt{g}\Phi+N^i\tmH_i+ (v_\Gamma+\Omega_0)\Gamma+v_\Omega
P_\Omega+v_N\pi_N) \ ,
\end{equation}
where
\begin{equation}
\Phi\equiv\frac{1}{\int d^3\bx\sqrt{g}}\mH_0+\Omega_0 \approx 0 \ ,
\end{equation}
and where we do not consider the modes $\bar{\Lambda}\approx 0
,\bar{p}_\Omega\approx 0$ that are canonically conjugate the second
class constraints that decouple from the theory. Before we proceed
further we should also modify the constraint $\bmH_T$ and $\tmH_i$
in such a way that  they Poisson commute with $\Gamma$ and $\Phi$.
In fact, let us consider following modification of the constraint
$\bmH_T$
\begin{equation}
\bmH'_T=\bmH_T+\frac{1}{32\pi Gg}g^{ij}\pi_{ij}\pi_N\ .
\end{equation}
Now it is easy to see that
\begin{equation}
\pb{\bmH'_T(\bx),\Gamma(\by)}=0 \ .
\end{equation}
Further we  have to  ensure that $\bmH'_T$ Poisson commute with the
constraint $\Phi$. Clearly we have $\pb{\bmH_T,H_0}\approx 0$,while
\begin{equation}
\pb{\bar{\bT}_T(N),\frac{1}{\int d^3\bx \sqrt{g}}}= 8\pi
G\left(\frac{1}{\int d^3\bx \sqrt{g}}\right)^2\int d^3\bx
\bar{N}\pi^{ij}g_{ij} \
\end{equation}
so that in order to cancel this contribution we extent the
constraint $\bmH'_T$ so that it has the form
\begin{equation}
\bmH_T''= \bmH'_T+8\pi G \left(\frac{1}{\int d^3\bx
\sqrt{g}}\right)^2
 \overline{\pi^{ij}g_{ij}}P_\Omega \ ,
\end{equation}
where by definition $\int d^3\bx
 \overline{\pi^{ij}g_{ij}}=0$.
 In the similar  way we modify the
diffeomorphism constraint $\tmH_i$ so that it Poisson commute with
$\Gamma$ (Note that it has vanishing Poisson bracket with $\Phi$ on
the constraint surface automatically)
%\begin{equation}
%\bmH'_i=\tmH_i+\pi_N\partial_i N
%\end{equation}
%since we have
%\begin{eqnarray}
%\pb{\bmH_T(\bx),\Gamma(\by)}=N\frac{1}{32\pi G}
%g^{ij}\pi_{ij}\delta(\bx-\by) -\frac{\sqrt{g}(\bx)}{\int d^3\bz
%\sqrt{g}}
% \nonumber \\
%\end{eqnarray}
%\begin{equation}
%\tmH'_i=\tmH_i+\pi^N \partial_i N
%\end{equation}
%so that we have
%\begin{eqnarray}
%\pb{\int d^3\bx N^i\tmH'_i,\Gamma}=
%-\partial_i N^i\sqrt{g}N-N^i\partial_i(\sqrt{g})N
%-\partial_i N N^i\sqrt{g}\approx -\partial_i N^i\nonumber \\
%\end{eqnarray}
\begin{eqnarray}
\bmH_i=\tmH_i+\partial_i\left[\frac{\pi_N}{\sqrt{g}}\right] \ .
%\bar{\mH}_i=\tmH'_i+\partial_i (fl \mH'_T) \ ,
%\nonumber \\
%\bar{\mH}_T=\tmH'_T+f' \pi_N \ ,
\nonumber \\
\end{eqnarray}
%First
%of all we have
%\begin{eqnarray}
%\pb{\bmH_T(\bx),\Gamma(\by)}= -\frac{1}{16\pi G}
%g^{ij}\mG_{ijkl}\pi^{kl}N-f'\sqrt{g}=
%\nonumber \\
%=\frac{1}{32\pi G}g_{ij}\pi^{ij}N-f'\sqrt{g}=0 \nonumber \\
%\end{eqnarray}
%and hence we see that we have to choose $f'$ to be equal to
%\begin{equation}
%f'=\frac{1}{32\pi G\sqrt{g}}g^{ij}\pi_{ij}N \ .
%\end{equation}
so that
\begin{eqnarray}
\pb{\bT_S(N^i),\Gamma(\by)}=
%-N^k\partial_k(N\sqrt{g})-
%\partial_i N^i N\sqrt{g}-\nonumber \\
%-\int d^3\bx \partial_iN^i[\frac{1}{\sqrt{g}}(-\sqrt{g})]= \nonumber \\
-N^k\partial_k \Gamma-\partial_i N^i\Gamma
%-\partial_i N^i+
%\partial_i N^i
\approx 0 \ .  \nonumber
\\
\end{eqnarray}
In summary we have following total Hamiltonian
\begin{equation}
H_T=\int d^3\bx (N\mH_T+\bar{v}_N\bar{\mH}''_T+v^N_0\int d^3\bx
\sqrt{g}\Phi +N^i\bmH_i+ (v_\Gamma+\Omega_0)\Gamma+v_\Omega
P_\Omega+v_N\pi_N) \
\end{equation}
and check stability of all constraints:
\begin{eqnarray}
\partial_t\pi_N=\pb{\pi_N,H_T}=-\mH_T-\Omega_0 \sqrt{g}-v_\Gamma
\sqrt{g} \approx -v_\Gamma\sqrt{g}=0 \nonumber \\
\end{eqnarray}
that   implies that $v_\Gamma=0$. For $P_\Omega $ we obtain
\begin{equation}
\partial_t P_\Omega=\pb{P_\Omega,H_T}=-\Gamma-v^N_0\int d^3\bx\sqrt{g}=0
\end{equation}
that determines $v^N_{0}$ to be equal to zero. For the constraint
$\Gamma$ we find
\begin{equation}
\partial_t \Gamma=\pb{\Gamma,H_T}=v_N=0
\end{equation}
and we find $v_N=0$. Finally for $\Phi$ we obtain
\begin{equation}
\partial_t \Phi=\pb{\Phi, H_T}=
v_\Omega=0
\end{equation}
and we again find $v_\Omega=0$. Now we should proceed to the
analysis of the time evolution of the constraints $\bmH_i,\bmH''_T$.
However these constraints Poisson commute with the second class
constraints by construction and also the Poisson brackets among
themselves  vanish on the constraint surface according to
(\ref{pbbTS}) and (\ref{pbbTSbar}).

 In summary we found that
$P_\Omega\approx 0, \pi_N\approx 0 ,\Gamma\approx 0, \Phi\approx 0$
are the second class constraints. Solving these constraints we
eliminate $N,\pi_N,\Omega_0,P_\Omega$ as functions of dynamical
variables. Then the remaining constraints  $\bmH''_T,\bmH_i$ form
the set of  $4\infty^3-1$ first class constraints with agreement
with \cite{Kuchar:1991xd}.

\section{Unimodular Gravity in Henneaux-Teitelboim Form}\label{third}
In this section we  consider the Henneaux-Teitelboim formulation of
unimodular gravity that is based on the existence of  the space-time
vector density $\mF^\mu$. In this case the  action has the form
\cite{Henneaux:1989zc}
\begin{equation}
S=\frac{1}{16\pi G}\int d^4x
[\sqrt{-\hg}({}^{(4)}R-2\Lambda)+2\Lambda
\partial_\mu \mF^\mu] \ ,
\end{equation}
where $\Lambda(\bx,t)$ is space-time dependent Lagrange
multiplicator. Our goal is to perform the canonical analysis of
given theory. Firstly we find following collection of the primary
constraints
\begin{equation}
\pi_N\approx 0 \ , \quad  \pi_i\approx 0 \ , \quad  \Gamma\equiv
p_t^{\mF}-\frac{1}{8\pi G}\Lambda \approx 0 \ , \quad p^{\mF}_i\approx 0 \
, \quad  p_\Lambda \approx 0 \ ,
\end{equation}
where $p^\mF_t,p^\mF_i$ are momenta conjugate to $\mF^t,\mF^i$
respectively with following canonical Poisson brackets
\begin{equation}
\pb{\mF^t(\bx),p^\mF_t(\by)}=\delta(\bx-\by) \ ,
\pb{\mF^i(\bx),p^\mF_j(\by)}=\delta_i^j\delta(\bx-\by) \ .
\end{equation}
Then we again find that the bare Hamiltonian with primary
constraints included has the form
\begin{eqnarray}
H=\int d^3\bx \left(N\left(\mH_T+p_t^\mF\sqrt{g}\right)
+N^i\mH_i-\frac{1}{8\pi G}\Lambda\partial_i\mF^i+v_N\pi_N+v^i\pi_i+
v_\Lambda p_\Lambda+u_i p_\mF^i+u_\Gamma \Gamma \right) \ , \nonumber \\
\end{eqnarray}
where with the help of  the constraint $\Gamma$ we replaced
$\frac{1}{8\pi G}\sqrt{g}\Lambda$ with $p_t^\mF\sqrt{g}$.
 Now requirement of the
preservation of the primary constraints imply following secondary
constraints
\begin{eqnarray}
\partial_t\pi_N&=&\pb{\pi_N,H_T}=-
(\mH_T+p_t^\mF\sqrt{g})\equiv -\mH_T'\approx 0 \ ,
\nonumber \\
\partial_t\pi_i&=&\pb{\pi_i,H_T}=-\mH_i\approx 0 \ , \nonumber \\
\partial_t \Gamma&=&\pb{\Gamma,H_T}=-\frac{v_\Lambda}{8\pi G}=0 \ ,
\nonumber \\
\partial_t p_i^\mF&=&\pb{p_i^\mF,H_T}=\frac{1}{8\pi G}\partial_i
\Lambda \ , \nonumber \\
\partial_t p_\Lambda&=&\pb{p_\Lambda,H_T}=
\frac{1}{8\pi G}\partial_i \mF^i+\frac{1}{8\pi G}u_\Gamma=0 \ .\nonumber \\
\end{eqnarray}
The third and the fifth equation determines the Lagrange multipliers
$v_\Lambda$ and $u_\Gamma$. As in previous section we find that the
fourth equation implies that $\bar{\Lambda}(t,\bx)=0$ while the zero
mode part $\Lambda_0$ is not determined.
%In case of the variable $\Lambda$ it is natural to perform the split
%of this variable as follows
%\begin{equation}
%\Lambda(\bx,t)=\Lambda_0(t)+\bar{\Lambda}(\bx,t) \ , \quad
%\Lambda_0(t)=\frac{1}{\int d^3\bx\sqrt{g}}\int d^3\bx \sqrt{g}
%\Lambda(t,\bx)
%\end{equation}
% $\int d^3\bx \sqrt{g}\bar{\Lambda}(t,\bx)=0$.
%Then the constraints $\partial_i\Lambda=0$ is equivalent to the
%condition
%\begin{equation}
%\bar{\Lambda}(t,\bx)=K
%\end{equation}
%that by definition of $\bar{\Lambda}(t,\bx)$ we have
%\begin{equation}
%\int d^3\bx \bar{\Lambda}(\bx,t)=0 \Rightarrow K=0
%\end{equation}
In other words we have the second class constraints $\bar{\Lambda}=
0 \ , p_{\bar{\Lambda}}=0$ so we will not consider these modes
anywhere and restrict ourselves to the case of the zero mode of $\Lambda$.
%Now with the help of the constraint $\Gamma$ we express $\mH'_T$ as
%\begin{equation}
%\mH'_T=\mH_T+p_\mF^t \ , \quad  \pb{\mH'_T,p_\Gamma}=0 \ .
%\end{equation}
%Note that $\mF$ is tensor density so that we determine its transf.
%properties as follows. We write the interaction term as
%\begin{equation}
%\int d^4x \partial_\mu \Lambda \mF^\mu
%\end{equation}
%and hence
%\begin{eqnarray}
%\int d^4x' \partial_{\mu'} \Lambda' \mF'^\mu(x')=
%\nonumber \\
%=\int d^4x |\frac{\partial x'^ \mu}{\partial x^\nu}|
%\partial_\nu \Lambda \frac{\partial x^\nu}{\partial x'^\mu}
%f\frac{\partial x'^\mu}{\partial x^\nu}\mF^\nu f=\nonumber \\
%=\int d^4x |\frac{\partial x'^ \mu}{\partial x^\nu}| f\partial_\mu \Lambda
%\mF^\mu \nonumber \\
%\end{eqnarray}
%so that in order to be given expression the same we derive
%following transformation property of the tensor density
%\begin{equation}
%\mF'^\mu(x')=\frac{1}{|\frac{\partial x'^ \mu}{\partial x^\nu}|}
%\mF^\nu(x)\frac{\partial x'^\mu}{\partial x^\nu} \ .
%\end{equation}
%or equivalently
%\begin{equation}
%\delta \mF^\mu=\mF'^\mu(x)-\mF^\mu=
%-\epsilon^\nu\partial_\nu\mF^\mu-\partial_\nu\epsilon^\nu \mF^\mu
%-\mF^\nu\partial_\nu \epsilon^\mu \ .
%\end{equation}
%Further, using the fact that $\mF^i$ disappear from the Hamiltonian
%using the integration by part we can gauge away $\mF^i$
%since $p_i^{\mF}$ is the first class constraint.
 Finally we modify
$\mH_i$ in order to incorporate the transformation rule for $\Lambda$
\begin{equation}
\mH'_i=\mH_i+p_\Lambda\partial_i \Lambda \
\end{equation} so
that the total Hamiltonian has the form
\begin{equation}
H_T=\int d^3\bx (N\mH'_T+N^i\mH'_i+v_\Lambda p_\Lambda +u_\Gamma
\Gamma) \ ,
\end{equation}
where we also used integration by parts that eliminates the term
$\Lambda\partial_i\mF^i$.
 Finally we see that
$p_\Lambda$ and $\Gamma$ are the second class constraint so that we
can eliminate $p_\Lambda$ and $\Lambda$ from the theory. As a result
we find the theory with  $4\infty^3$ the  first class constraints
$\mH'_T,\mH'_i$ for the dynamical variables
$g_{ij},\pi^{ij},p_t^\mF,\mF^t$. Note that the Hamiltonian does not
depend on $\mF^t$ explicitly and hence we see that $p_t^\mF$ is
constant on-shell. In other words $p_t^\mF$ plays the role of the
cosmological constant which however is not included into the theory
by hand but it arises as a consequence of the dynamics of the
unimodular theory in Henneaux-Teitelboim formulation.

\vskip .5in \noindent {\bf Acknowledgement:}
\\
  This work   was
supported by the Grant agency of the Czech republic under the grant
P201/12/G028.

\end{document}